\documentstyle[twocolumn,aps,epsfig,floats]{revtex}
\draft
\begin{document}
%==============================================================================
\twocolumn[\hsize\textwidth\columnwidth\hsize\csname @twocolumnfalse\endcsname

\title{Electronic structures of doped anatase $\rm TiO_{2}$:
       $\rm Ti_{1-x}M_{x}O_{2}$ (M=Co, Mn, Fe, Ni)} 

\author {Min Sik Park, S. K. Kwon, and B. I. Min}

\address{Department of Physics and electron Spin Science Center,  \\
Pohang University of Science and
	Technology, Pohang 790-784, Korea}

\date{\today}
\maketitle

\begin{abstract}

We have investigated electronic structures of a room temperature
diluted magnetic semiconductor: Co-doped anatase $\rm TiO_{2}$.
We have obtained the half-metallic ground state in the local-spin-density
approximation (LSDA) but the insulating ground state in the LSDA+$U$+SO 
incorporating the spin-orbit interaction.
In the stoichiometric case, the low spin state of Co is realized
with the substantially large orbital moment. 
However, in the presence of oxygen vacancies near Co, the spin state of Co 
becomes intermediate. The ferromagnetisms in the metallic and insulating 
phases are accounted for by the double-exchange-like and the
superexchange mechanism, respectively.
Further, the magnetic ground states are obtained for Mn and Fe doped
$\rm TiO_{2}$, while the paramagnetic ground state for Ni-doped $\rm TiO_{2}$.

\end{abstract} 

]
%]
%\narrowtext
%==============================================================================

%\section{Introduction}
%\label{sec:intro}
   
Diluted magnetic semiconductors (DMSs) have been studied extensively
for last decades, because of their potential usages of both charge and 
spin degrees of freedom of carriers in the electronic devices, namely 
the spintronics.
There have been trials based on two types of DMS families:
II-VI such as Mn-doped CdTe and ZnSe\cite{Furdyna}, 
and III-V such as Mn-doped GaAs\cite{Ohno}. 
Especially, the latter attracts great attention,
because it becomes a ferromagnetic (FM) DMS having the
Curie temperature $T_{C} \sim 110K$. 
Motivated by the above FM DMS, recent research effort has been 
focused on developing new FM semiconductors operating at
room temperature\cite{Dietl,Sato}. It has been reported that 
the FM DMSs are realized in other types of systems too
\cite{Medvedkin,Matsumoto,Ando}.

Matsumoto {\it et al.}\cite{Matsumoto} fabricated Co-doped anatase
$\rm TiO_{2}$ thin film samples, $\rm Ti_{1-x}Co_{x}O_{2}$, 
using the combinatorial pulsed-laser-deposition (PLD)
molecular beam epitaxy (MBE) technique. 
A sizable amount of Co, up to $x =0.08$, is soluble in anatase $\rm TiO_{2}$.
Using the scanning SQUID microscope, they observed  
the magnetic domain structures in Co-doped films
characteristic of the FM long range ordering.
The measured saturated magnetic moment per Co ion was $0.32\mu_{B}$ 
apparently in the low spin state and $T_{C}$ was estimated 
to be higher than 400K. This sample is conductive at room temperature,
but becomes semiconducting at low temperature.
It also exhibits a large positive magnetoresistance of $60\%$ at 2K 
in a field of 8T.
Due to transparent property of the system, it can be used in integrated 
circuits and storage devices with display units\cite{Ohno1}. 

More recently the Co-doped anatase
$\rm TiO_{2}$ film grown by the oxygen-plasma-assisted (OPA) MBE was
reported by Chambers {\it et al.}\cite{Chambers}. 
They claimed that magnetic properties 
of the OPA-MBE grown material are better than those of the PLD-MBE grown material
because considerably larger saturated magnetic moment of $1.26\mu_{B}$/Co
is observed, which seems to be consistent better with the 
low spin state of Co. 
The unquenched orbital moment of Co in the asymmetric crystalline field 
was ascribed to the enhanced magnetic moment.
The Co $L$-edge x-ray absorption (XAS) 
spectrum of Co-doped $\rm TiO_{2}$ is similar to that of CoTiO$_3$,
whereby the formal oxidation state of Co$^{2+}$ has been suggested.  
They also found that the
magnetic and structural properties depend critically on the Co 
distribution which varies widely with the growth condition.

Hence the magnetic properties of Co-doped anatase $\rm TiO_{2}$
are still controversial.
To explore these properties, the essential first step is
to study the electronic structure of Co-doped anatase $\rm TiO_{2}$.
In this study, we have investigated 
electronic structures of Co-doped anatase $\rm TiO_{2}$: 
$\rm Ti_{1-x}Co_{x}O_{2}$ ($x=0.0625$ and 0.125) using the 
linearized muffin-tin orbital (LMTO) band method both in the 
local-spin-density approximation 
(LSDA) and the LSDA+$U$+SO incorporating the Coulomb correlation
interaction $U$ and the spin-orbit interaction \cite{Kwon}. 
For a comparison, we have also  investigated electronic structures of 
other transition metal doped $\rm TiO_{2}$: $\rm Ti_{1-x}M_{x}O_{2}$
(M=Mn, Fe, Ni).

$\rm TiO_{2}$ has three kinds of structures, rutile, anatase,
and brookite. The space group of anatase structure is tetragonal $I4_{1}/amd$.
The anatase $\rm TiO_{2}$ is composed of 
stacked edge-sharing octahedrons formed by six O anions.
Ti atoms are in the interstitial sites of octahedrons
that are distorted with different bond lengths 
between the apical (1.979 $\AA$) and 
the equatorial (1.932 $\AA$) Ti-O bond and with the Ti-O-Ti angle 156.3$^{o}$.
For $\rm Ti_{1-x}Co_{x}O_{2}$ $(x=0.0625)$, we have considered 
a supercell containing sixteen formula units in the primitive unit cell
by replacing one Ti by Co ($\rm Ti_{15}Co_{1}O_{32}$: 
$a=b=7.570, c=9.514$ \AA).
Sixteen empty spheres are employed in the interstitial sites to enhance 
the packing ratio for the LMTO band calculation.

%-----------------------------------------
\begin{figure}[t]
\epsfig{file=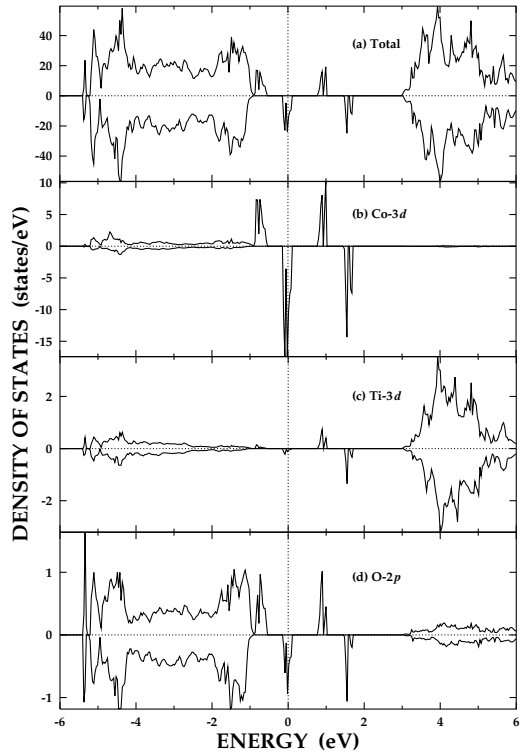,width=7.10cm}
\caption{The LSDA total and PLDOS
         of $\rm Ti_{1-x}Co_{x}O_{2}$ $(x=0.0625)$.
}
\label{Colsda}
\end{figure}
%-----------------------------------------

%\section{Results and Discussion}
%\label{sec:results}

We have first calculated the electronic structure of anatase $\rm TiO_{2}$
without doping elements.
The overall band structure of
the present LMTO result is consistent with existing results\cite{Asahi},
except that the energy gap is estimated a bit larger, $\sim 4$ eV,
as compared to the FLAPW result of $\sim 2$ eV \cite{Asahi}. 
Although the present result is closer to the experimental energy gap 
of 3.2 eV, it is likely that the energy gap is overestimated 
due to the open structure of anatase 
$\rm TiO_{2}$ and the minimal basis of the LMTO band method. 
The valence band top and the conduction band bottom correspond to mainly
O-$2p$ and Ti-$3d$ states, respectively.
The electronic transport and magnetic experiments on anatase $\rm TiO_{2}$
indicate peculiar properties, such as shallow donor level and high mobility of 
the n-type carriers 
due to intrinsic oxygen off-stoichiometry \cite{Forro}.
 
To examine the energetics of $\rm Ti_{1-x}Co_{x}O_{2}$
between the FM and antiferromagnetic (AFM) 
configurations of Co ions, we have performed the LSDA 
band calculation for anatase $\rm Ti_{1-x}Co_{x}O_{2}$
($x=0.125$).
In this case, there are two Co ions in the unit cell 
($\rm Ti_{14}Co_{2}O_{32}$) separated by
5.353\AA. Inbetween two Co ions, there are one Ti and two O ions.  
As a result, we have obtained that the FM phase 
has {\it half-metallic} electronic structure, while the AFM 
phase has semiconducting electronic structure.
Total energies are very close, but the FM phase is 
lower by $\sim 6 m$Ry than the AFM phase.
Hence, in the following discussion, we will consider only the FM
configurations of Co ions.

Now, we have performed band calculations for
anatase $\rm Ti_{1-x}Co_{x}O_{2}$ ($x=0.0625$).
Figure~\ref{Colsda} shows the density of states (DOS)
obtained from the LSDA band calculation.
The energy gap between O-$2p$ and Ti-$3d$ states
is almost unchanged by Co doping and most of Co $d$ states are located in 
the energy gap region.
Noteworthy is the {\it half-metallic} nature in this system,
reminiscent of the Mn-doped GaAs\cite{shirai,Akai,JHPark}, 
that is, the conduction electrons at the Fermi level 
$\rm E_F$ are $100 \%$ spin-polarized.
The carrier types, however, are different between two.
Here the Fermi level cuts the Co $t_{2g}$ states,
whereas, in Mn-doped GaAs, the Fermi level cuts mainly the As-$p$ states
since Mn-$3d$ states are located far below $\rm E_F$.
The different carrier types would give rise to the different magnetic 
mechanisms as discussed below\cite{Kacman}.
The crystal field splitting between $t_{2g}$ 
and $e_{g}$ state is larger than the exchange splitting between $t_{2g}$ 
states, suggesting the low spin state of Co.
The total spin magnetic moment is $1\mu_{B}$, which comes mostly from Co ions.
Ignoring the extended Co-$d$ states between $-5$ and $-1$ eV,
which are hybridized bonding states with O-$2p$ states,
the characters of localized $d$ states are mainly
$t_{2g}^{3}$ spin-up and $t_{2g}^{2}$ spin-down states,
seemingly corresponding to the ionic valence of Co$^{4+}$.
However, almost two electrons are occupied in
the extended Co $d$ states,
and so the total occupancy of $d$ states amounts to $d^7$.

%-----------------------------------------
%-----------------------------------------
\begin{figure}[t]
\epsfig{file=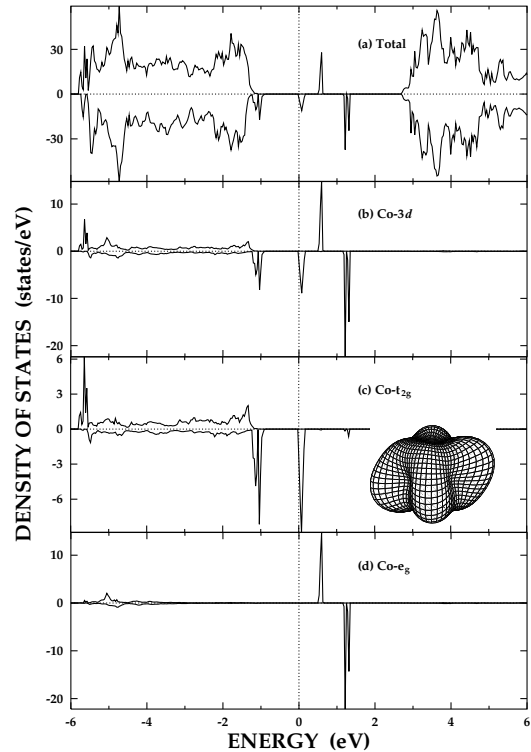,width=7.10cm}
\caption{The LSDA+$U$+SO total and PLDOS
	 for $\rm Ti_{1-x}Co_{x}O_{2}$$(x=0.0625)$.
         The angular distribution of occupied Co-3$d$ 
	spin-down states (inset (c)).
}
\label{Co}
\end{figure}
%-----------------------------------------

The half-metallic LSDA result for Co-doped $\rm TiO_{2}$ seems
to be compatible with the metallic resistivity  behavior above 100K.
Further, in view of the carrier type of Co-$3d$,
the FM ground state can be understood based on
the {\it double-exchange-like} mechanism, {\it e.g.}, 
the kinetic energy gain through 
the hopping of fully spin-polarized carriers in the half-metallic system.
This is contrary to the case of Mn-doped GaAs in which As-$p$ hole carriers
mediate the {\it RKKY-like} exchange interaction.
Note, however, that at low temperature the system behaves 
as an insulator \cite{Matsumoto}.
Since the unfilled $t_{2g}$ states near $\rm E_F$  are very narrow, 
one can expect that the Coulomb correlation interaction
and/or the Jahn-Teller interaction would induce the
metal-insulator transition. The Jahn-Teller effect would be relatively
weak because the relevant orbitals near $\rm E_F$ are $t_{2g}$ states.
Thus we have explored the effect of the Coulomb correlation interaction
using the LSDA+$U$+SO band method. The spin-orbit interaction is taken into 
account to describe properly atomic-like Co-$t_{2g}$ states.

Indeed, the LSDA+$U$+SO band calculation with parameter values of $U=3.0$ eV 
and $J=0.87$ eV for Co-3$d$ electrons yields the semiconducting ground state 
in accord with the experiment.
The DOS plot in Fig.~\ref{Co} shows that 
the $t_{2g}$ spin-down states are separated by the $U$ effect with
the band gap size of $\sim 0.8$ eV.
The total spin magnetic moment of $1\mu_{B}$ and the occupancy of $d^7$
are also obtained by the LSDA+$U$+SO.
In the inset of Fig.~\ref{Co}(c), the angular distribution of 
occupied Co $3d$ spin-down states are plotted, based on the orbital 
occupancies. The main contribution to these states comes from
the $d_{xy}$ state mixed partially with $d_{yz}$ and $d_{zx}$, 
which is reflected in the shape of the angular 
distribution. Due to the spin-orbit effect, however,
the shape is a bit asymmetric and distorted.

Evidently, atomic-like Co $t_{2g}$ states would yield the unquenched orbital 
moment. In fact, Co-ion has remarkably large orbital magnetic 
moment of 0.9 $\mu_{B}$ which is as large as that in CoO ($\sim 1.0 \mu_{B}$)
\cite{Kwon2}.
The large orbital moment arises from occupied $t_{2g}$ spin-down
states split by the Coulomb correlation and the spin-orbit 
interaction (Fig.~\ref{Co}).
The orbital moment is polarized in parallel with the spin moment, 
and so the total magnetic moment amounts to 1.9 $\mu_{B}$/Co.
The large orbital magnetic moment is in agreement with the 
expectation
by Chambers {\it et al.}\cite{Chambers}, but the total magnetic moment 
1.9 $\mu_{B}$ is much larger than their experimental value 1.26 $\mu_{B}$.

%-----------------------------------------
\begin{figure}[t]
\epsfig{file=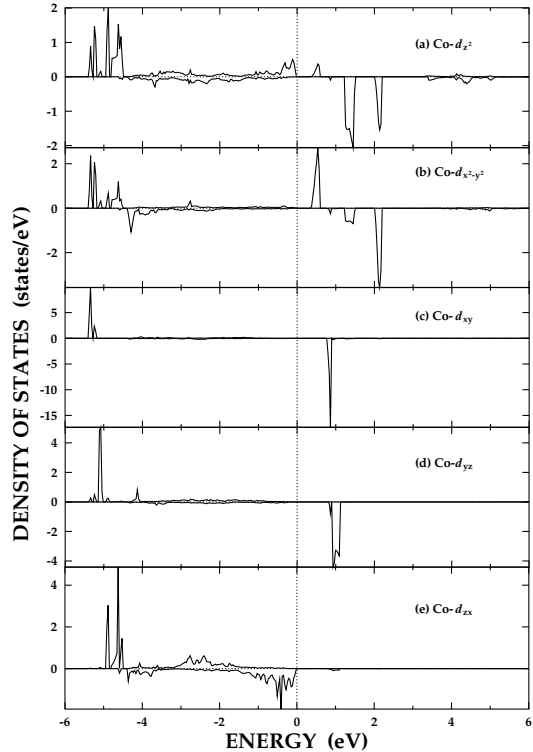,width=7.10cm}
\caption{The LSDA+$U$+SO Co-PLDOS of oxygen deficient Co-doped
         $\rm TiO_{2}$ $\rm (Ti_{15}Co_{1}O_{31})$.
}
\label{odef}
\end{figure}
%-----------------------------------------
As mentioned above, oxygen vacancies are easily formed in the
anatase $\rm TiO_{2}$, and so there will also be intrinsic oxygen vacancies
in Co-doped $\rm TiO_{2}$. We thus examined the effects of
oxygen deficiency in the Co-doped TiO$_{2}$.
By removing one oxygen atom in the supercell ($\rm Ti_{15}Co_{1}O_{31}$), 
the formal valence of Co becomes Co$^{2+}$ in the ionic picture.
We have considered two cases of removing an oxygen atom:
(i) from the Ti-contained octahedron, and (ii) from the Co-contained 
octahedron.
For the oxygen vacancy near the Ti site, essentially the same Co-$3d$
projected local density of states (PLDOS)
are obtained as for the stoichiometric case with the low spin 1.0 $\mu_{B}$/Co 
and the orbital magnetic moment 0.9 $\mu_{B}$/Co,
implying that the Co sites are not affected much by the vacancy \cite{Lsda}.
On the other hand, for the oxygen vacancy near the Co site, very different 
features are revealed: the intermediate (close to the high) spin state 
is realized rather than the low spin state with the spin
magnetic moment of 2.53 $\mu_{B}$/Co.
The intermediate spin state in this case results from the
reduced crystal field in the pyramidal structure composed of five 
oxygen anions. 
Then the $d_{z^2}$ and $d_{zx}$ states become more stabilized than in
the case of stoichiometric octahedral structure (see Fig.~\ref{odef}).
The configuration of occupied states becomes
$d^7$ with spin-up $t_{2g}^3$$d_{z^2}^1$ and extended
$d_{x^2-y^2}^{1-\alpha}$ bonding states and
spin-down $e_{g}^1$ and extended $t_{2g}^{1+\alpha}$ (mainly $d_{zx}$) 
bonding states.  Due to decreased $t_{2g}$ characters near $\rm E_F$,
the orbital magnetic moment is reduced to 0.28 $\mu_{B}$.
Comparing the total energies between above two cases,
the oxygen vacancy near the Ti site
is more stable than the oxygen vacancy near the Co 
site. Therefore oxygen vacancies tend to be formed 
mostly in the Ti-contained octahedrons without affecting the Co spin state.
However, the possible oxygen vacancies near Co-sites formed during
the non-equilibrium MBE growth would influence drastically  
the magnetic properties in Co-doped TiO$_{2}$ film. 
This explains the observation that the
magnetic properties depend critically on the film growth condition.

\begin{figure}[t]
\epsfig{file=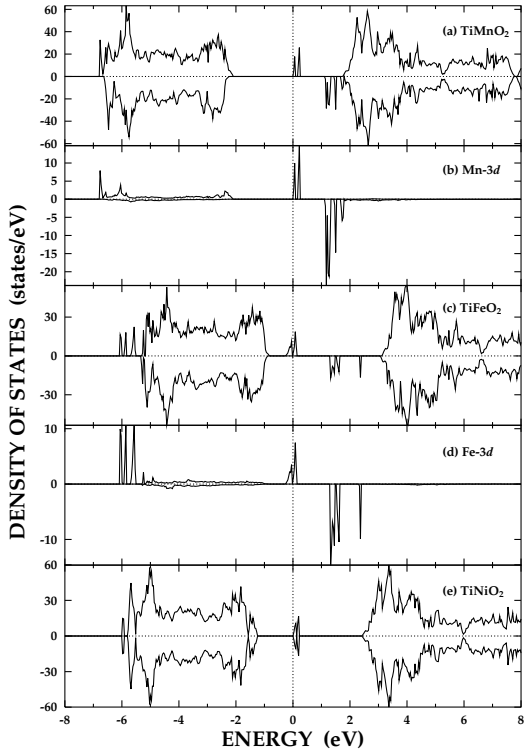,width=7.10cm}
\caption{The LSDA+$U$ DOS of Mn, Fe-doped DOS, and the LSDA DOS of Ni-doped 
	 $\rm TiO_{2}$$(x=0.0625)$.
}
\label{FeMn}
\end{figure}

For comparison, we have also examined electronic structures of 
other transition metal doped anatase TiO$_{2}$:
$\rm Ti_{1-x}M_{x}O_{2}$ (M=Mn, Fe, Ni) for x=0.0625.
The same $U$ and $J$ parameters were employed for 
Mn and Fe $3d$ electrons.
Figure~\ref{FeMn} shows that Ni-doped TiO$_{2}$ has the paramagnetic ground
state, whereas, Mn and Fe doped systems have the magnetic ground states  
with local magnetic moments of 3.0 and 3.7 $ \mu_{B}$, respectively. 
Ignoring the extended bonding $d$ states,
the apparent nominal valences look like Mn$^{4+}$ ($d^3$) 
and Fe$^{4+}$ ($d^4$). Including the extended states, however,
the electron configuration for the Mn-doped case becomes
$d^5$ with spin-up $t_{2g}^{3}e_{g}^{1}$ states and spin-down
bonding $t_{2g}^{\alpha}e_{g}^{1-\alpha}$ states.
Likewise, for the Fe-doped case,
the electron configuration becomes $d^6$ with
spin-up $t_{2g}^{3}e_{g}^{2-\alpha}$ states and 
spin-down bonding $t_{2g}^{\beta}e_{g}^{1+\alpha-\beta}$ states.
Therefore, for both cases, the intermediate close to the
high spin states are realized even without oxygen defects.

%-----------------------------------------
%-----------------------------------------

No evidence of FM behavior
has been observed in Mn or Fe-doped TiO$_{2}$ film yet.
The different magnetic natures in these systems 
are presumably due to their different electronic structures.
Note that the characters of unoccupied states near $\rm E_F$ are different:
$t_{2g}$ for the  Co-doped TiO$_{2}$ (Fig.~\ref{Co}), whereas $e_{g}$ for the
Mn and Fe-doped TiO$_{2}$.
For Mn-doped case, even the LSDA yields the insulating electronic structure, 
and so, by considering only the localized states,
the superexchange via occupied 
Mn($t_{2g}$)-O($p_{\sigma}$)-Mn($t_{2g}$) orbitals would lead to 
the nearest neighbor AFM interaction.
On the other hand, for Fe-doped case, the LSDA yields the
half-metallic electronic structure, as for Co-doped case.
The Jahn-Teller effects would be more operative in this case
because of the $e_{g}$ characters near $\rm E_F$, 
which will drive the structural distortion and the concomitant metal-insulator 
transition.  Then the AFM phase is more likely to be stabilized.
In contrast, the Co-doped TiO$_{2}$ even in its insulating phase
would have the nearest neighbor FM superexchange interaction
via Co($t_{2g}$)-O($p_{\pi}$)-Co($t_{2g}$) kinetic-exchange energy gain.
Therefore, the FM phases in the metallic and insulating 
states are accounted for by the double-exchange-like and the
superexchange mechanism, respectively.
In reality, however, the situation may not be that simple, 
since there could be some effects of extra-carriers originating from 
the intrinsic or extrinsic O-vacancies. 
The present study serves to provide basic band structure informations 
for understanding the magnetic mechanisms in these systems.

In conclusion,
the LSDA yields the {\it half-metallic} ground state for Co-doped 
$\rm TiO_{2}$ with the carrier type of mainly Co $3d$ states.
In contrast, the LSDA+$U$+SO yields the insulating ground state
and the low spin state with $1 \mu_{B}$ spin moment 
and $0.9 \mu_{B}$ orbital moment per Co ion.  
The possible oxygen vacancies near Co sites substantially affect 
the magnetic properties: the intermediate spin state of Co 
with 2.53 $\mu_{B}$ spin moment is realized and the 
orbital moment is reduced to 0.28 $\mu_{B}$.
We have also found that Mn and Fe-doped  $\rm TiO_{2}$ have 
the magnetic ground states, 
while Ni-doped $\rm TiO_{2}$ has the paramagnetic ground state.

Acknowledgments $-$
This work was supported by the KOSEF through the eSSC at POSTECH
and in part by the BK21 Project.
Helpful discussions with J.-H. Park and Y.H. Jeong are greatly appreciated.


\begin{references}

   \bibitem{Furdyna} J. K. Furdyna and J. Kossut, {\it DMSs},
           {\bf 25} of Semiconductor and Semimetals
           Academic Press, New York, (1988). 
   \bibitem{Ohno} H. Ohno {\it et al.},
        Appl. Phys. Lett. {\bf 69}, 363 (1996).
   \bibitem{Dietl} T. Dietl {\it et al.},
	Science {\bf 287}, 1019 (2000).
   \bibitem{Sato} K. Sato and H. K. Yoshida,
        Jpn. J. Appl. Phys. {\bf 39}, L555 (2000).
   \bibitem{Medvedkin} G. A. Medvedkin {\it et al.},
	Jpn. J. Appl. Phys. {\bf 39}, L949 (2000).
   \bibitem{Matsumoto} Y. Matsumoto {\it et al.},
	Science {\bf 291}, 854 (2001).
   \bibitem{Ando} K. Ando {\it et al.},
        Appl. Phys. Lett. {\bf 78}, 2700 (2001);
	K. Ueda, H. Tabata, and T. Kawai, {\it ibid} {\bf 79}, 988 (2001).
   \bibitem{Ohno1} H. Ohno, Science. {\bf 291}, 840 (2001).
   \bibitem{Chambers} S. A. Chambers {\it et al.},
	Appl. Phys. Lett {\bf 79}, 3467 (2001).
   \bibitem{Kwon} S. K. Kwon and B. I. Min, 
        Phys. Rev. Lett. {\bf 84}, 3970 (2000).
   \bibitem{Asahi} R. Asahi {\it et al.}, 
	Phys. Rev. B {\bf 61}, 7459 (2000); and references therein.
   \bibitem{Forro} L. Forro {\it et al.},
	J. Appl. Phys. {\bf 75}, 633 (1994).
%   \bibitem{Chauvet} O. Chauvet {\it et al.},
%        Solid State Commun. {\bf 93}, 667 (1995).
   \bibitem{shirai} M. Shirai {\it et al.},
	J. Magn. Magn. Mater. {\bf 177-181}, 1383 (1998). 
   \bibitem{Akai} H. Akai, Phys. Rev. Lett. {\bf 81}, 3002 (1998).
   \bibitem{JHPark} J. H. Park, S. K. Kwon, B. I. Min,
        Physica B {\bf 281-282}, 703 (2000).
   \bibitem{Kacman} P. Kacman, 
	Semicond. Sci. Technol. {\bf 16}, 25 (2001).
   \bibitem{Kwon2} S. K. Kwon and B. I. Min, 
	Phys. Rev. B {\bf 62}, 73 (2000).
   \bibitem{Lsda} For the O-vacancy near Ti, vacancy induced impurity states 
	appear just below the conduction band, but the Co PLDOS is 
	essentially the same. Half-metallic electronic structures are obtained 
	in the LSDA for both cases of O-vacancies.



\end{references}
\end{document}